\documentclass[compsoc,conference,a4paper,10pt,times]{IEEEtran}
\usepackage{cite}
\usepackage{amsmath,amssymb,amsfonts}
\usepackage{algorithmic}
\usepackage{graphicx}
\usepackage{textcomp}
\usepackage{bmpsize}
\usepackage{xcolor}
\usepackage{lipsum}
\usepackage{array}
\usepackage{ragged2e}
\newcolumntype{P}[1]{>{\RaggedRight\hspace{0pt}}p{#1}}
\usepackage[colorlinks=true,urlcolor=black]{hyperref}
\def\BibTeX{{\rm B\kern-.05em{\sc i\kern-.025em b}\kern-.08em
    T\kern-.1667em\lower.7ex\hbox{E}\kern-.125emX}}
\begin{document}

\title{``Security should be there by default'': Investigating how journalists \\perceive and respond to risks from the Internet of Things}

\author{\IEEEauthorblockN{Anjuli R. K. Shere}
\IEEEauthorblockA{\textit{Department of Computer Science} \\
\textit{University of Oxford, UK}\\
anjuli.shere@new.ox.ac.uk}
\and
\IEEEauthorblockN{Jason R. C. Nurse}
\IEEEauthorblockA{\textit{School of Computing} \\
\textit{University of Kent, UK}\\
j.r.c.nurse@kent.ac.uk}
\and
\IEEEauthorblockN{Ivan Flechais}
\IEEEauthorblockA{\textit{Department of Computer Science} \\
\textit{University of Oxford, UK}\\
ivan.flechais@cs.ox.ac.uk}
}

\maketitle

\begin{abstract}
Journalists have long been the targets of both physical and cyber-attacks from well-resourced adversaries. Internet of Things (IoT) devices are arguably a new avenue of threat towards journalists through both targeted and generalised cyber-physical exploitation. This study comprises three parts: First, we interviewed 11 journalists and surveyed 5 further journalists, to determine the extent to which journalists perceive threats through the IoT, particularly via consumer IoT devices. Second, we surveyed 34 cyber security experts to establish if and how lay-people can combat IoT threats. Third, we compared these findings to assess journalists' knowledge of threats, and whether their protective mechanisms would be effective against experts' depictions and predictions of IoT threats. Our results indicate that journalists generally are unaware of IoT-related risks and are not adequately protecting themselves; this considers cases where they possess IoT devices, or where they enter IoT-enabled environments (e.g., at work or home). Expert recommendations spanned both immediate and long-term mitigation methods, including practical actions that are technical and socio-political in nature. However, all proposed individual mitigation methods are likely to be short-term solutions, with 26 of 34 (76.5\%) of cyber security experts responding that within the next five years it will not be possible for the public to opt-out of interaction with the IoT.
\end{abstract}


\section{Introduction and Background}
The seemingly unlimited growth of the global technology industry is matched by a concerning expansion of users' attack surfaces. Mass manufacturing of hardware and software that can be rushed to market to meet the demand for cheap products has resulted in a deluge of technologies that are intended to increase user convenience, without maintaining a minimum standard of security. In recent years, the threat landscape for even members of the general public has been changing rapidly, particularly with the mass proliferation of so-called ``smart-'' devices forming an ``Internet of Things'' (IoT)~\cite{forbes16}. More formally, an IoT can be regarded as ``a network that connects uniquely identifiable `Things' to the Internet. The `Things' have sensing/actuation and potential programmability capabilities. Through the exploitation of unique identification and sensing, information about the `Thing' can be collected and the state of the `Thing' can be changed from anywhere, anytime, by anything''~\cite{minerva15}.

Common examples of novel consumer IoT devices include voice assistants (e.g., Amazon Echo, Google Home) and wearables (e.g., fitness trackers and smart watches). The consumer IoT offers various benefits, but the combination of common vulnerabilities and the generation of personal user data and even bystander information, both in terms of content and metadata, can pose threats to privacy and security. In 2019, the BBC aired a documentary in recognition of the importance of educating consumers about the risks inherent in available gadgets~\cite{BBC19}. These were often marketed as intending to improve user security but were, in fact, demonstrably easily compromised to manipulate device actions and mine user information. Various organisations have acknowledged the risks of the IoT, including the UK government, who provided mitigation guidance for users~\cite{DCMS18}, and has recently proposed laws for manufacturers to enshrine a Code of Practice that would make such guidance feasible~\cite{BBC20}. Researchers have also explored this domain in great depth highlighting a series of complex risks and solutions~\cite{alladi2020consumer,hsu2016empirical}.

While these measures may be effective against indiscriminate hackers, they are unlikely to protect individuals who are targeted, particularly by well-resourced attackers. Journalists have long considered how best to protect themselves, their sources (i.e., contacts who supply information for a news story), and their work from emerging and evolving threats. Traditional strategies for privacy and security are no longer effective against the ubiquity of IoT devices~\cite{hassan2019current,nurse2017security}. This is particularly given how they constantly collect, transmit and store data, including audio, visual and geolocation data that affect the physical and informational security of journalists and their sources. 

While journalists are likely aware of traditional risks (e.g., being tracked via a mobile phone's GPS), guidance documents by well-known supporting organisations such as the Rory Peck Trust~\cite{roryp19} and Reporters without Borders~\cite{unrwb17} overlook the potential for new IoT devices (e.g., wearables) to be used to collect (for example) scheduling, messages, geolocation and biometric data. The UNESCO and Reporters without Borders `Safety Guide for Journalists' warns that ``a smartphone can be treacherous''~\cite{unrwb17} but does not acknowledge other internet-connected devices. Already, instances of novel attacks on the press include those conducted through botnets (which can be IoT-enabled). Examples include Distributed Denial of Service (DDoS) attacks on news organisations~\cite{ellao19} and the National Union of Journalists in the Philippines~\cite{mindanews19}, seemingly intended to censor the press by limiting the availability and accessibility of their work. 

In the research community, there are studies (e.g., ~\cite{mcgregor2015investigating,mcgregor2016individual,mcgregor2017weakest,henrichsen2019breaking}) that have assessed the threats journalists face from conventional cyber-threats to phones, laptops and communications channels. For instance, attacks have been discussed on communication channels, personal devices and email, which target journalists, their sources and employers. To address these, a range of strategies have been proposed involving using encryption, secure or/and anonymous communication, multifactor authentication, and protecting device metadata. What is lacking, however, is research into the threats journalists face from IoT devices, particularly those that they may use, or be exposed to. This is seriously needed given the difficulty in assessing risk in, and developing ways to secure, these devices~\cite{nurse2017security}.

As the IoT features more in journalists' daily lives (both at home through consumer devices, and at work in smart environments), researching its implications are imperative. The goal of this study therefore is to evaluate how journalists perceive and respond to the risks present in this ubiquitous technology platform. Moreover, where relevant, our work also provides an updated capture of the risks and risk treatment behaviours with existing technologies; thereby adding directly to work such as~\cite{mcgregor2015investigating}. 

Our objectives are threefold: first, through interviewing and surveying members of the media industry, we intend to examine the extent to which journalists understand the risks inherent in the IoT (including how the IoT might exacerbate pre-existing threats, such as security issues regarding remote communications methods) and to collect information on how they attempt to combat perceived IoT-related threats. Second, we survey cyber security experts to confirm how, if at all, lay-people (``the general public'') can effectively protect their privacy and security against threats through the IoT. Third, we seek to compare and contrast the self-disclosed security behaviours of the journalists with the expert recommendations, in order to find out whether the journalists should maintain or could improve their security strategies to protect against both immediate and long-term IoT threats. We hope this study can provide a unique insight into an understudied area of work and motivate future contributions on the topic.

\section{Research Questions and Methodology}

To achieve the objectives of our research, we have outlined the following research questions:
\begin{enumerate}
    \item To what extent do journalists perceive threats from the Internet of Things (IoT)?
    \item What protective tools and methods do journalists feel are currently effective in increasing their cyber security against all perceived cyber-threats, including any from the IoT? 
    \item Are the protective measures reported as in use by journalists actually effective, when compared to the known recommendations from contemporary literature and experts in cyber security?
\end{enumerate}

\subsection{Interviews with Journalists}
Based on the first two research questions, an interview guide for participating journalist  was formulated. Potential interviewees were originally selected based on their published work. In particular we looked for interviewees that demonstrated an interest in, or understanding of, cyber security or other high-risk topics which could make them targets of cyber-attacks by well-resourced adversaries. After selection, these individuals were contacted and asked if they would be interested in participating in the study. This work applied a semi-structured interview process to guide the questioning but also to allow flexibility.

Our questions were premised on the belief that having a heightened awareness of the risks attached to researching high profile topics is likely to cause journalists to adopt defensive strategies earlier than others in the industry. Therefore, candidates were recruited initially as a purposive sample of our professional contacts and the pool of respondents was subsequently expanded through snowball sampling. All interviewees were chosen for their likelihood of being among those who should be best versed in the dangers of the IoT that are specific to journalism, with the objective, if possible, of recording and proliferating their protective procedures to guide other journalists.

All interviews were transcribed, and then analysed using qualitative content analysis. We followed a 5-step process~\cite{erlingsson2017hands} for abstracting qualitative interview data to recognise and group themes across interviews. The steps were: (1) noting the ``meaning unit'' (i.e., direct quote); (2) condensing this in our own words; (3) creating a relevant and brief ``code'' to label the quote; (4) grouping quotes with similar codes into the same category to demonstrate their factual relevance to each other in context; and (5) clustering categories into overarching themes. These themes would then form the basis for the primary analysis and discussions. For these interviews and the other data gathering methods below, ethical approval was granted by the institutional Research Ethics Committee.

\subsection{Online Survey with Journalists}
An online survey was developed in collaboration with the Association for International Broadcasting (AIB) for journalists who declined an interview. The survey questions were formulated based on the interview questions. Although questions mainly gathered quantitative data, the questionnaire also included space for long-form comments. The survey was sent by the AIB to a number of their journalist contacts, regardless of demonstrated interest in cyber security or the IoT, because of the prevalence of the IoT and its propensity for collecting substantial amounts of sensitive data. As such, the questionnaire is intended to discover what journalists are broadly doing to protect themselves against third-party exploitation (be it from criminals, governments, or intelligence services) of IoT (or, smart) devices. We analysed the data gathered using descriptive statistics, and qualitative content analysis similar to the interviews.

\subsection{Online Survey with Cyber Security Experts}
We also developed an online questionnaire aimed at cyber security experts. This was intended to support in answering the third research question, i.e., to evaluate the protective tools and techniques used by journalists against the IoT threats to which even members of the general public would likely be exposed. All cyber security expert respondents were either invited via academic mailing lists or via the researchers' social media posts, which requested participants who were involved in cyber security. 

The expert survey aimed to record three aspects: digital and physical threats associated with the use of smart devices, whether and how they would advise members of the public to cope with these threats, and the extent to which members of the general public can opt-out of the use of IoT devices now and in the future. We asked experts for recommendations for the general public, rather than for journalists specifically, because we wanted to ensure that expert responses did not overlook basic threat models that would affect everyone. This was intended to establish a baseline of knowledge about both possible IoT threats and proposed protective tactics more generally. Six open-ended questions in the questionnaire collected expert recommendations to inform individuals on how to minimise or mitigate IoT threats.

\section{Results}
\label{sec:results}
This section details the findings of the study. It begins by introducing the professionals who participated and their demographic information. We then present the range of IoT threats and threat vectors that journalists face; thus answering the first research question. This is followed by an assessment and explanation of how journalists have responded to these threats with protection strategies; this allows us to address research question two.  Lastly, we compare journalists' protective measures with the recommendations outlined by experts as it pertains to cyber security practices for the IoT (for research question three).

\subsection{Participants}
A total of 11 individuals positively responded to our request for interview participation out of the 36 contacted. Interviewees reported having a variety of the specialisms, including cyber security, drug trafficking, conflicts and defence, human rights and transnational organised crime. All of these specialisms were considered high-risk in terms of targeting by highly skilled state and criminal actors. It should be noted that one participant (I10) was not a journalist, but instead works closely with journalists and news organisations on cyber security training and response. Given this expertise and experience, we decided it would be advantageous to interview this person as well. On average, the interviews lasted for 45 minutes. 

There were five respondents to the online journalist survey; each with varied specialisms and experiences. Table~\ref{tab:interviewees} and Table~\ref{tab:AIB} provide basic demographic information of interviewees and survey respondents respectively. For our cyber security expert questionnaire, a total of 34 experts were recruited. Of these, 29 (85.3\%) claimed to have conducted research in the field of cyber security.

\begin{table}[htbp]
\caption{Details of interviewees}
\begin{tabular}{|l|P{0.22\textwidth}|P{0.1\textwidth}|}
\hline
\textbf{Interviewee} & \textbf{Role(s) held in the media sector} & \textbf{Years working as a, or with, journalist(s)} \\ \hline
I1 &  Journalist & 6 \\ \hline
I2 &  Journalist & 15 \\ \hline
I3 &  Journalist & 20 \\ \hline
I4 &  Journalist & 10 \\ \hline
I5 &  Journalist / media executive & 40+ \\ \hline
I6 & Journalist / foreign \newline correspondent & 21 \\ \hline
I7 & Journalist / academic researcher & 25 \\ \hline
I8 & Journalist / foreign \newline correspondent & 4 \\ \hline
I9 & Journalist / academic researcher / cyber security advisor and trainer & 25+ \\ \hline
I10 & Cyber security advisor & 5+ \\ \hline
I11 & Journalist / foreign \newline correspondent / consultant & 37 \\ \hline
\end{tabular}
\label{tab:interviewees}
\end{table}

\begin{table}[htbp]
\caption{Details of AIB survey respondents}
\begin{tabular}{|l|P{0.08\textwidth}|P{0.06\textwidth}|P{0.06\textwidth}|P{0.065\textwidth}|}
\hline
\textbf{Respondent} & \textbf{Journalist specialism} & \textbf{Cyber security is a subject element of work} & \textbf{Has been the target of cyber attacks} & \textbf{Years working as a journalist} \\ \hline
S1 & Live reporting & Yes & Yes & 38 \\ \hline
S2 & Technology & Yes & Yes & 29 \\ \hline
S3 & None & Yes & Yes & 25 \\ \hline
S4 & International refugee crisis & No & No & 10 \\ \hline
S5 & Broadcast journalism & No & No & 10 \\ \hline
\end{tabular}
\label{tab:AIB}
\end{table}

Due to the high-risk nature of some participants' work, particularly within the interviewees, more demographic information is not divulged. Additionally, in the following analysis of the study's results, the gender-neutral pronoun ``they'' has been used to further obfuscate identities. In the upcoming sections, journalist participants will be referred to with pseudonymised identifiers beginning with I (interviewees, i.e., I1-I11) and S (journalists surveyed through the questionnaire disseminated by the AIB, i.e., S1-S5), as per Tables~\ref{tab:interviewees} and~\ref{tab:AIB}.

\subsection{IoT Threats that Journalists Face}

\subsubsection{Threats to Information}
IoT-related dangers to journalists can manifest in different ways. We regard threats to information as threats to the confidentiality, integrity and availability of raw or analysed information.

Two interviewees, both of whom carry out work advising and training journalists on security (I9, I10), stated that their experience indicated that most journalists with whom they interact are unaware of IoT-related threat models. I10, a cyber security advisor, responded that journalists and their organisations were reactionary and tended to seek help only after being attacked. I10 also stated that many journalists, ``\textit{don't understand the risks of their basic devices so don't see how adding more internet-connected devices expands the attack surface}'', and that the threat was not only to information but in some cases, safety. By two interviewees' own admission (I4, I7), additional risk emanated from reckless behaviour. Even though aware of the importance of using hardened professional networks and sanitised devices, they confessed to taking work calls near to potentially insecure smart-home entertainment systems. This therefore could threaten the confidentiality of sensitive data on their work, or their sources.

\begin{figure}[htbp]
\centerline{\includegraphics[width=1\columnwidth]{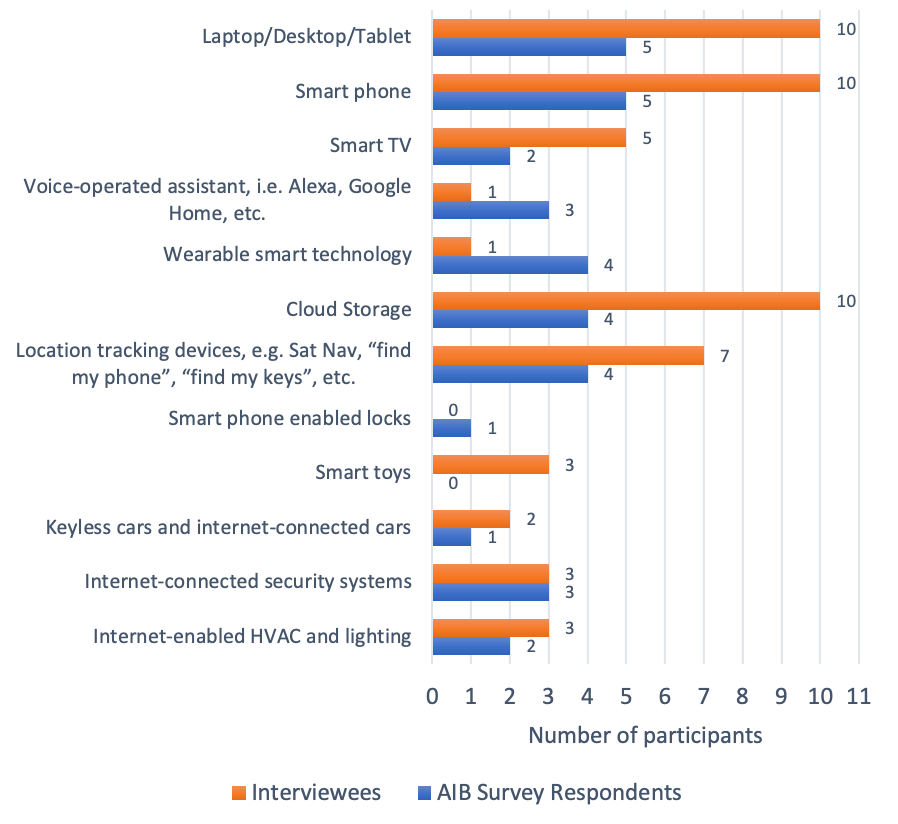}}
\caption{Devices that interviewees and AIB survey respondents use.}
\label{fig:devices}
\end{figure}

IoT utility often relies on the same function that poses a threat to journalists, such as the need for voice-activated devices to be constantly listening~\cite{williams2017privacy}. Only two interviewees (I1, I4) explicitly mentioned or alluded to the knowledge that IoT cloud databases (for instance) were points of vulnerability, either for information loss, or because they could divulge locational metadata from self-tracking services. This was surprising given the number of journalists who use cloud services, as shown in Figure~\ref{fig:devices}; one interviewee chose not to answer these questions. Various smart technologies are also already in use (e.g., wearables, smart TVs and location tracking devices), even in this set of journalists who engage in high-risk topics. Another key point here is that at some stage it may well be  impossible for journalists to remain constantly vigilant against ambient IoT surveillance. 

One AIB survey respondent, S1, stated that they have a strategy of, ``\textit{keeping sensitive data away from the devices and if necessary away from the cloud}''. Their answer inferred a mental separation of IoT devices and the cloud, also signifying a lack of awareness of how most IoT devices work. That is, IoT devices and the cloud are coupled; this allows the resource-constrained IoT device to focus on data collection and presentation, while the cloud is used for more resource-heavy tasks such data storage and processing~\cite{williams2017privacy}. It was common for interviewees to emphasise abstractly their awareness that newer IoT devices are often less secure and do not use strong encryption, while neglecting the technical details of other IoT-specific vulnerabilities, such as cloud-reliant architectures.

Three AIB survey respondents (S2, S4 and S5) explicitly expressed their concern that IoT devices pose a threat to anonymity of sources. However, S5 also noted that their work is reliant on, ``\textit{these internet enabled mediums}'', indicating the complexity of the choice to limit their access to such technologies. 

There were a few threat-aware journalists worth highlighting. For instance, the concept of ``\textit{a porous border between the internet and the tangible world}'' frightened I4, as they likened IoT devices to windows in a house: ``\textit{My assumption is always that the more devices you have, the more entry-points for attackers, just like if there are more windows in your house there are more entry-points for burglars}''. This signals an awareness of how the increased prevalence of IoT devices expands the attack surface of an individual and their information environment.

I4 also highlighted a significant issue with devices and keeping them, and the data they hold, secure. They gave the example of video game consoles to illustrate their lived experience of manufacturers being financially motivated and so abandoning devices (i.e., not releasing patches and updates, or making additional features incompatible with previous models). This therefore acts to make the devices redundant, and eventually insecure; a noteworthy threat given the pace at which IoT devices are being released and abandoned.

\subsubsection{Legal Threats}
Legal threats to journalists occur when states implement legislation that grants them intrusive or sweeping powers. Regarding the IoT and the free press, these often manifest as data retention laws, national security loopholes in data protection regulations, the criminalisation of whistleblower and journalistic activities, or bans on security provisions such as strong encryption.

Ten interviewees admitted or intimated that they are concerned that IoT devices amass data that can be accessed by authoritarian or overreaching democratic governments with a track record of coercing companies into divulging this data in the name of national security (all except I1). Some of these interviewees (I3, I4, I6) explicitly mentioned concerns about the risks inherent in poorly secured IoT devices because of the likelihood of state surveillance of journalists and a history of attempts to use overreaching legal mechanisms to conduct leak enquiries. 

A perception that most standard IoT devices lack end-to-end encryption was also the subject of four responses (I2, I4, I5, I10) regarding legal threats to journalists. Three participants (I2, I7, I8) also expressed fears regarding the lack of accountability employed by governments to justify data retention and anti-encryption laws (such as those in Australia~\cite{theguard19}), and six (I2, I3, I5, I6, I7, I8) indicated concern about legal loopholes in data protection laws (such as the ``Legitimate Interest'' justification in the General Data Protection Regulation (GDPR)).

I7 noted that their research has unearthed clear worries in the journalistic communities in countries such as Australia. In this case, the concern is linked to Australia's ban on strong encryption. In an IoT world, this may mean that access to the huge quantities of IoT data and metadata can be gathered and then used by governments to retroactively build cases against a targeted individual. The result of this could be prosecution, incarceration or social alienation such that a journalist would not be able to continue to effectively operate in their profession.

\subsubsection{Physical Threats}
Given the cyber-physical nature of the IoT, IoT threats could be escalated to threats to the physical security, well-being and safety of journalists and their sources. I3, who consented to being identified as a Mexican journalist with experience covering drug trafficking stories, stated that our interview was the first time that they had discussed cyber security because, ``\textit{it's not linked enough in countries like mine to survival}''. Their answer elucidated that, traditionally, the perception of cyber-threats is focused on information, rather than being linked to potential physical attacks. 

While it is true that technologies like smart phones, laptops and tablets are primarily at risk of loss of data pertaining to a news piece, novel IoT environments (where there exist fitness trackers, smart offices, smart light bulbs and internet-connected vehicles) arguably represent a new kind of cyber-physical system that has more urgent implications for both the data and physical well-being of individuals~\cite{privint19}. I7, who has conducted academic research into the interface of cyber security and journalism, referred to the belief of free speech and regulation expert, Dr. Katherine Gelber, that the prevalence of the IoT comes with global dependence on the IoT for, ``\textit{production, dissemination, the basic tenets of our jobs, so there's no way to separate the kinds of attacks that affect information from those that affect us physically}''. 

Indeed, I3 noted experiences that had abstractly concerned them, disclosing that, ``\textit{sometimes the TV switches on by itself and that's not normal}'', and going on to express that ``\textit{everyone is watching you, the web is like Big Brother}''. This is an example of a plausible scenario in which manipulation of the physical features of an IoT device stimulate a psychological reaction (fear and stress) in a journalist, potentially causing subsequent physical issues---from health responses to erratic or antisocial behaviour that further isolates a journalist.

The asymmetrical nature of the threat means that it is difficult to anticipate threat levels and to know how to prioritise different defence and source protection strategies. Interviewee I9, who has a role as a cyber security trainer for journalists, spoke of the commonly held belief that cyber-threats against journalists come primarily from states. Their work has revealed that this is often the case, but also that journalists are frequently subject to cyber-attacks from the private sector, particularly from banks, which are highly underestimated in terms of the time and money they spend hunting whistleblowers through surveillance of journalists. Due to the potent global influence of these public and private institutions, the interviewee argued that there is a higher likelihood of targeted cyber-physical attacks on journalists through the IoT that result in clandestine consequences. I2 also noted that attribution of threats is difficult for exactly this reason. 

I7 argued that appeals to national security frameworks to allow states to access IoT contents and metadata could initiate a slippery slope towards ``\textit{sedition and treason}'', for which the ramifications are potentially physical and life-threatening. I10, whose primary role is as a cyber security advisor, noted that, ``\textit{in our work it's very crucial that we understand that digital security is just one component. There are also physical and legal threats and other kinds of attack. It's pretty common in our work that we see these be interconnected. Even with the best protections in the world, if the weakest link in the chain is physical or legal security, the threat model is too weak}''.

I11 also emphasised the real-world, physical impact of cyber threats. They noted that, ``\textit{there are so many unpredictables within the expanding network, specifically because of IoT, that there is an elevated risk of something happening in the cyber sphere having a very significant impact in the real world}''. I9 pointed out that the normalisation of an increased prevalence of IoT devices could also lead to populations becoming dependent on their continued function, which would give a zero-day attack the power to destabilise societies and the global order in both kinetic and non-kinetic ways. I10, who also assists journalists with cyber-attack recovery, further stated that, ``\textit{there are a lot of differences between the capabilities of the adversaries and the journalists being attacked [and] for someone who doesn't have a network to support them, it's very hard to fight those battles}''.

Regarding targeted attacks on individual journalists, I2 acknowledged that the proliferation of IoT sensors in urban spaces, both formally and informally creating so-called `smart cities', would aid adversaries in tracking and locating journalists, thereby facilitating planned opportunities for physical attacks. This could, of course, also pose a threat to information if devices were subsequently seized. On a smaller scale, I4 noted that they had fears, as both a consumer and a journalist, about the physical effects of the market for basic appliances (e.g., TVs and fridges) being flooded by IoT devices, thereby making non-internet-connected devices more expensive and less accessible and forcing IoT devices to be the new standard. They were concerned that an attack on their internet access and/or power supply would drastically affect the functionality of critical devices.

\subsubsection{Threat Vectors}
Our media participants (both interviewees and AIB survey respondents) also touched upon their understanding of IoT threat vectors, naming a variety of aspects of the IoT that are particularly susceptible to exploitation. We compared these to the answers from our 34 cyber security expert respondents and compiled the following two subsections based on all responses.

Cyber security expert survey respondents mentioned several potential attack vectors for IoT devices, most of which relate to inadequate security considerations of manufacturers and vendors. These included, ``\textit{reuse of code libraries}'' by developers, an inability to alter passwords, insecure communication protocols between devices and servers, a lack of patches and ``\textit{missing/delayed/inadequate firmware updates}'' post-release of products. These are all evidenced in current research~\cite{conti2018internet}.

These expert responses could be divided into two broad classes: privacy and access control. For access-control, their responses regarding threats could be classed as illicit access to information that indirectly facilitates physical harm, or altering the output of an IoT device so that it directly or indirectly facilitates or obfuscates physical harm to its environment and to its user. 

Both of these classes of threat could have particularly destructive consequences if enacted against journalists, whose role relies on trust---both that they can ensure source confidentiality, and that they can verify that their published work is truthful. Despite these considerations, when journalists' answers were compared to those from cyber security expert respondents, our media participants did not appear to have in-depth knowledge or understanding of threats or threat vectors. While all their concerns were realistic, we assessed that at least five of them (I1, I3, I4, I6, I7) lacked detail and were limited in scope. This suggests that journalists, at least based on our sample, do not have sufficient understanding of the multitude of potential threats that can be amplified because of the IoT.

\textbf{Privacy-related Threats}. Many interviewees expanded on two recurrent themes: that IoT devices are easily compromised as a result of (1) the financial incentive for manufacturers to externalise costs and insufficient industry standards and regulations (I1, I2, I4, I5, I9, I10, I11), and (2) systemic subscription to data capitalism (I1, I2, I4, I5). These points were corroborated by expert respondents, who noted that both the insecurity of IoT devices and their intentional data collection present threats to the public. One expert summed up both themes, saying ``\textit{there is a commercial incentive to build a badly set up panoptican [sic]}''.

Another privacy-related issue that was raised by three respondents (I1, I5, I8) linked to the masses of data accumulated and transferred by IoT devices: information leakage and theft resulting from weak or non-existent encryption and other poor data protection mechanisms, such as inadequate cloud or application security. Three AIB respondents (S3, S4 and S5) also explicitly mentioned encryption, while S1's answers did not mention data transfer and instead seemed to indicate that they try to physically and virtually isolate their data. 

Regarding strategies for the preservation of sources' integrity and confidentiality in light of the ever-expanding IoT market, S2 simply said, ``\textit{individual journalists do not have viable solutions for this}''. For journalists, unwanted disclosure of data could endanger their relationships with sources. Notably, one respondent (I7) mentioned the ``\textit{chilling effect}'' on news reporters as a potential outcome of increased awareness of the data collection and analytics resulting from expansion of the IoT.

\textbf{Access Control-related Threats}. Access control threats can have implications both in the cyber and physical realms. With respect to the latter, altering the output of an IoT device can both directly and indirectly cause physical harm to its environment and users and, one expert respondent noted, ``\textit{new emerging physical attack models [...] are likely to not be known to end users of these systems}''. Our study demonstrated that this is true of members of the media industry.

Interviewee I10 stated that often attribution is unnecessary after cyber-attacks on journalists, from the target's perspective, because they can logically link the attack to the subject of a particular story that they are planning to publish. Additionally, the constant presence of IoT technologies and the opacity of their settings would allow the creation of a dangerous environment even when the attacker is not physically present. Participants suggested that there are also scenarios in which cyber and physical attacks could be concurrent, such as, ``\textit{people using photos of valid users' faces to gain access to their smart locks}'' or tampering with burglar alarms, which could then lead to the purely physical threat of theft. These scenarios clearly demonstrate how attacks on the IoT could effectively hamper journalists' desire to work on stories involving highly resourced hostile adversaries.

One expert survey respondent pointed out that there is the greatest threat of physical harm to a number of people, ``\textit{where there is a convergence between IoT and [Operational Technology]/SCADA/[and Industrial Control Systems], [which] can result in denial of critical services or physical damage}''. Dependence on such systems ranges from individual to societal. Examples of the former include infrastructure devices (e.g., thermostats, gas and water level detectors and smoke alarms, all of which rely on accurate and timely sensor reporting), and even the use of embedded medical devices (e.g., pacemakers, which one respondent said, ``\textit{can easily be hacked}''). 

Other expert survey participants considered targeted attacks with immediate and overt ramifications, saying that, ``\textit{any device which can catch fire, explode, emit whatever, etc. will be hacked to do so}'', and giving more specific examples, such as, ``\textit{a heater overheating and causing a fire (or similar) or turbines being sped up until they self-destruct}''. It is clear that the prevalence of the IoT has the potential to facilitate massive targeting of individual journalists or suspected whistleblowers without likely attribution, which could have a rapid and effective negative effect on the capabilities of the free press.

\subsection{Protection Strategies against IoT Threats}

\subsubsection{Protective Tools and Methods that Journalists use to increase their Cyber Security}
When asked about the tools and techniques they employ to defend against IoT exploitation, every interviewee confirmed that their primary approach to risk mitigation was to minimise interactions with IoT devices and return to analogue methods of data collection, communication and storage. Above all, this recurrence of traditional journalistic habits was noted as being a crucial component of source protection. I7, however, acknowledged that it was regressive for newsgathering, as IoT technologies, ``\textit{would make the job easier and give us access to sources who were further away or evidence that is harder to find}''. Worryingly, perhaps because of a feeling of futility regarding the security of and protection from IoT devices, some respondents (I1, I3, I4) admitted that they did not have coherent or relevant security strategies. Participants I5 and I10, whose work involves managing or cyber security training numerous journalists, noted that their organisational risk assessment procedures do not take into account IoT threats.

\subsubsection{Physical Protection Strategies for Journalists}
All ten journalist interviewees (I1-I9, I11) were concerned that the spread of casual-use IoT devices could remove the confidentiality of source-journalist interactions. I10 was more attuned to journalists' vulnerabilities owing to their work as a civil society cyber security trainer. Respondents' techniques for minimising threats resulting from the IoT can be broadly classified as attempting to limit their access to the internet, and mainly focused on disconnecting devices and disabling applications when not in use. I9 also mentioned taking Faraday cage bags to meet sources so that, if batteries could not be removed, sources' devices could be isolated, allowing unintentional audio-recordings to be muffled and any signals to be blocked. 

Interviewees also mentioned using covers over cameras to physically limit the photo-taking functionality of IoT devices. However, all highlighted strategies were acknowledged to be temporary because of the dangers of remote access when internet-connection is built into the functionality of essential devices such as (smart) home security systems. Interviewees recognised that the main issue with physical strategies was the same as with all IoT threats; i.e., journalists cannot always control IoT presence in the environment where they meet with sources.

Resultantly, some journalists primarily rely on remote communications methods. However, as previously mentioned, interviewees (I4, I7) openly admitted to communicating important information in the presence of insecure IoT devices. This is likely to render any effort to use secure remote communications channels redundant, a problem that will only be exacerbated as the prevalence of ambient IoT technologies grows. 

While not specific to the IoT, there are a few related security practices worthy of note. For instance, nine interviewees discussed the challenges of finding and exclusively using secure channels for communication with sources. Two interviewees (I2 and I11) detailed their use of specific encrypted messaging platforms, such as Signal and hardened emails, like ProtonMail. Others simply mentioned PGP encryption or that they try to rely on encrypted communication mechanisms more abstractly. However, I5 suggested that it was hard to verify whether these methods of securing remote communications were sufficient, and thus also discussed attempts not to put any potentially source-identifying information into a digital format. I5 further expressed that this, ``\textit{has made it more difficult to communicate}''. Acknowledging the same challenges, I8 was nevertheless aware that completely avoiding remote communications channels is unrealistic even if encryption methods are not fully satisfactory. This highlights a series of security usability challenges with traditional devices that can also apply to the IoT.

Many IoT devices now have inbuilt email functionality (e.g., smart watches and even smart fridges), providing an additional pivot-point for attackers to access email correspondence, thereby further lowering confidence in email security, confidentiality and integrity. Interviewees differed in their approaches to email security. I11 claimed that the core of their security regime is to, ``\textit{assume that what you're doing is public communications}'', while I1's sources only communicate with them using burner accounts and then the emails are bounced through multiple accounts, ``\textit{so the [audit] trail is muddled}''. 

Additionally, much of the reliance on, or avoidance of, emails depends on source location and context. For example, I2 preferred not to communicate via email if the story is sensitive while, at organisations for which I7 has worked, ``\textit{some stories have been shut down completely if there have been emails}'' because the channel is ``\textit{not considered hygienic and that's often enough to end a source-relationship or to kill a story altogether, so it's a form of self-censorship}''. As a range of IoT devices now either have access to, or can send, emails, these are of direct concern to journalists' work.

\subsubsection{Allocation of Responsibility for Journalists' Cyber Security}
Repeated themes from interview respondents included that, ``\textit{the responsibility to educate journalists about IoT threats lies with news organisations}'' (I7) and that often, ``\textit{the threats come from governments so we wouldn't have incentive to listen to their advice}'' (I7). Multiple respondents (most of whom preferred to not have their responses even attributed post-pseudonymisation) also believed that states and state-affiliated agencies could not be trusted to inform journalists about IoT threats, as they were responsible for most instances of targeted exploitation of journalists' devices. As a result, proposed regulation of the IoT industry or reforms of data-related legislation would likely be met with scepticism by any journalists who understand technological threats. 
Instead, news organisations, journalists' unions and NGOs were the suggested parties for education and training of journalists against IoT threats. 

Additionally, many interviewees were sceptical about involving technology companies in the responsibility of protection of journalists as they felt that the technology industry tended to be more aligned with state interests. However, others, such as I10, felt that, ``\textit{the first line should be security experts, developers, making guidelines so the consumer knows that their device has basic security settings embedded}''. They pointed out that, ``\textit{the real question is whether it's reasonable to ask a journalist in distress in an oppressive regime to do all these checks. Security should be there by default}''. The implication from this interviewee's perspective is that there is a need for the development and legislation of standards for the IoT technology industry. 

Furthermore, journalists are increasingly freelancing or working for underfunded fledgling organisations that depend on the savvy of individual employees or on the advice of NGOs. Even I5, a media executive, said, ``\textit{I try to follow good and balanced advice from professional organisations for journalists, like the Electronic Frontier Foundation}''. The EFF is well known for its focus on privacy and provision of guidance and tools.

I2 and I6 both mentioned that the information that informs how they maximise their own security or brief sources on potential threats was collated by the security teams of organisations for which they work. I10 argued that it is, ``\textit{a challenge to update the risk assessment [used by their organisation, which deals specifically with advising actors such as journalists on cyber-threats] because every day more IoT devices are coming out and even the ones you think you should trust have stories of being a risk}''. They noted that because of the huge increase in internet-connected devices released, even in communities that should be more conscious of IoT threats, risk assessments are clearly lagging behind.

\subsection{Comparing Journalists' Protective Strategies with Security Recommendations from Experts}
Of our 34 expert respondents, 20 (58.8\%) felt that members of the public could not currently opt out of the use of the IoT in some way. Moreover, 26 (76.5\%) felt that within five years the public---including journalists---would find it almost impossible to opt out of having their information vulnerable to IoT devices, even if they avoided personal ownership of them. This point is critical for our work as it directly invalidates a primary protective measure suggested by all interviewees; i.e., the core of their (often informal) IoT-related security strategies is to avoid ownership of, and restrict contact with, IoT devices. This approach is simply not feasible. 

Experts suggested few solutions: in the short-term individuals could focus on efforts to avoid, subvert, or limit the ability of IoT devices to collect personal information. These include changing default settings to make passwords stronger, activating device firewalls, avoiding public Wi-Fi, segmenting networks so that critical and non-critical IoT devices could not interact, and purposefully inputting incorrect user information in order to impede accurate data collection. One expert said, ``\textit{physical and digital controls of such devices might offer some protection against privacy/security threats}'', which supports the utility of some mitigations implemented by journalist respondents, such as the use of camera covers and mechanisms to obstruct microphones like Faraday bags.

In the longer-term, expert respondents recommended better education about the potential threats of IoT devices (30.8\% of respondents), lobbying governments and private companies to create industry standards around consumer safety related to IoT devices (26.9\%), and a focus on the need for united public activism in order to push for protective legislation and education (11.5\%) was viewed as vital. This links to the pivotal point previously touched upon by I10: the free press is currently staving off near-constant attacks from highly skilled and well-resourced adversaries. Therefore, how can society expect individual journalists to implement both fundamental technical and high-level cyber security measures against evolving emerging threats emanating from the IoT? As we have mentioned, when asked who should be responsible for journalistic cyber security, media participants largely endorsed an institutional approach. This suggests that they appreciate, at least in part, the theme of most expert responses: that IoT threats are systemic and can only be combated in the long-term through structural shifts that require coordinated efforts.

\section{Discussion}
\label{sec:discussion}

\subsection{The Pervasive IoT Threat to Journalists}
The cyber-physical nature of the IoT has facilitated a myriad of novel anticipatory threat models, which can target the journalist, their devices, or their sources. As we have seen, use of IoT devices---either knowingly or unknowingly---can result in the compromise of information or sources, and even physical harm. The networked nature of the IoT means that gaining unauthorised access to one device can allow an attacker to compromise an entire smart environment. 
This factor is exacerbated by the general poor security of IoT devices~\cite{hassan2019current}, which means that users cannot consistently anticipate or control device behaviour, including data collection and sharing. Crucially for journalists, whose jobs rely on public trust, this also threatens to call into question data integrity as there is a risk that data could be compromised with, both on a device and while in transit, which would threaten the ability to trust that source material is unedited. 

Additionally, the availability of data held by multiple devices can be compromised through DoS attacks targeting even one IoT node in the network. This could mean active censorship, with journalists being locked out of smart devices or laptops, and denied access to information. This point becomes even more crucial when we consider that there have already been efforts that seek to fully incorporate the IoT in journalism as can be seen in the 2019 Journalism of Things Conference~\cite{vicari19jot}. Potentially most significantly, unauthorised access to IoT devices can result in host takeovers and their co-option into a botnet, which could then be used to enact DDoS attacks against other, better secured targets, such as entire media organisations. We covered examples of similar targeted attacks earlier in this article~\cite{ellao19,mindanews19}.

Fear of consequences for speaking freely can cause journalists to self-censor as IoT sensors are integrated into almost all environments, effectively creating a panopticon that can identify, track, record and analyse personal data on a global scale. This is the well-documented ``\textit{chilling effect}''~\cite{mills2016reluctant} of constant surveillance on free speech, enabled by technological advancement and an erosion of legislative safeguards for protection of individual privacy and a free press. 

Threats to the physical safety and well-being of journalists due to the IoT were a primary concern for study participants. In terms of conducting massive disruption or destruction operations through IoT exploitation, the respondents with the most acute understanding of potential physical threats to journalists through IoT devices (who all had experience in both journalism and research into journalists' security) were also best able to envision threat models for large-scale kinetic attacks. These attacks could have a generalised effect or could target people likely to be in a specific place at a specific time, which could arguably be used against multiple members of the media at once. 

At a more personal level, IoT devices can be targeted as platform of attack on individual journalists. For instance, wearables or smart appliances could have social media access~\cite{theguar19teen,nurse2015smart}, which then may be abused with disproportionate effects for journalists if hackers were to publicly share false or inflammatory information that could cause them to be discredited and lead to physical violence. Regarding altering the output of an IoT device so that the device directly causes physical harm to its environment and to its users, there are several threat models, both current and anticipatory. These range from crashing autonomous vehicles to taking over smart light bulbs. The latter could have a number of effects on the cyber and physical realms, from clandestine user data exfiltration~\cite{maiti2019light,nurse2015smart}, to their use to trigger epileptic attacks in photosensitive users~\cite{boingb16}. 

Potential scenarios in which IoT devices can be hijacked are likely to be difficult to corroborate, as Conti et al.'s research has demonstrated, digital forensics is difficult in many IoT devices. And, more specifically, ``\textit{detecting presence of IoT systems is quite a challenge considering these devices are designed to work passively and autonomously [...] in most cases when an IoT device is identified there is no documented method or a reliable tool to collect residual evidences from the device in a forensically sound manner}''~\cite{conti2018internet}. This highlights the significance of the challenge and pervasiveness of the threat. Another key factor here is the difficulty in assessing threats and risks in the IoT generally; this is a problem that is still arguably a unresolved research and practice~\cite{nurse2017security}. 

Lastly, there is the danger of normalisation of a reliance on the remote-access functionality of many IoT technologies, such as the ability to use one's phone to confirm that smart locks are engaged, which could then be exploited or malfunction. This could cause property to be physically less secure, as well as affecting users' mental and emotional well-being by suddenly removing a safety blanket upon which they had been dependent.

\subsection{The Need for IoT Risk Awareness and Protection Strategies for Journalists}
There is a clear need for better strategies to inform and protect journalists against the increasing number and variety of IoT threats. According to participants, both increasing awareness of the threats (many of which we presented earlier) and supporting protection approaches (via tools and best practice guidance) are crucial. Key stakeholders that could play a central role in  informing these strategies include news organisations (thus, employers), journalists' unions, and NGOs. 

Security-aware journalists can also be an invaluable source of guidance, and indeed, this seems to be the way that many security tips are currently being shared amongst peers. We also saw a number of IoT security and privacy recommendations provided by experts; including changing default settings, segmenting networks and better security education. Once defined, appropriate security and privacy guidance could be disseminated within well-read publications such as the UNESCO and Reporters Without Borders `Safety Guide for Journalists', or the Digital Risk Assessment guide by the Rory Peck Trust. The EFF may also be a suitable outlet given it is already being used by journalists.  This would ensure that guidance on how to address IoT threats is widely accessible.

It was noteworthy, though not surprising given the participant sample, that most respondents were openly scathing about the possibility of relying on governments to improve IoT security and protect journalists from related threats. This is even as some states, such as the UK~\cite{dcms19sbd} and France~\cite{etsi19iot}, have already attempted a push to legislate for security by design in IoT devices. 

The reason for this scepticism is likely due to the reality that in certain countries, legal provisions can be used by government agencies and state-affiliated organisations to unmask sources and monitor journalists without the need to justify actions to the public. There is a disparity between states' largely unfettered access to data and the amount of information that journalists can glean from governments; this is exemplified clearly by the 42\% UK government approval rate of Freedom of Interest requests in 2018~\cite{theecon19}. Access to the vast amounts of data collected and conveyed by IoT networks could allow corrupt government bodies and law enforcement agencies to manipulate information to incriminate journalists and censor news stories. This reality highlights the need for other key stakeholders---even those within academia---to step forward and create, and maintain, IoT risk awareness and protection strategies for journalists.

\subsection{Limitations}

There are some limitations that should be noted when considering the wider implications of our work. Firstly, we engaged with a constrained sample of individuals, both within our set of journalists and experts. Moreover, most of the participating journalists were involved with cyber security or other high-risk topics which could make them targets of physical or cyber-attacks by well-resourced adversaries. This may therefore mean that the threats we outline are not \textit{currently} appropriate for all journalists using IoT devices. This may well change in the future as IoT devices become more embedded in society.

Secondly, in our cyber security expert survey, we asked experts about IoT risks and protection strategies for the general public, rather than for journalists specifically. This was done in an attempt to prevent experts from overlooking basic threat models or protection approaches that would pertain to anyone; journalists are, after all, also individuals who interact with consumer technologies. We do, however, acknowledge that not specifically asking expert respondents about IoT risks to journalists may have influenced their responses towards underestimation of risk. Another approach could have been to ask experts about journalists specifically, or to instead, adopt a staged process to expert engagement, where threat perceptions and protection approaches from journalists fed directly into the design of the expert survey. This could potentially provide a more robust basis for comparison.

\section{Conclusion and Future work}
IoT proliferation across homes, workplaces and social spaces will drastically expand the attack surface surrounding all individuals but will be particularly dangerous to communities that are already highly targeted. In this work, we determined that one such community, journalists, does not accurately understand IoT-related threats and are not adequately protecting themselves. 

We did this by: first, interviewing and surveying a total of 16 members of the media regarding their perceptions of IoT threats to journalists. Second, surveying 34 cyber security experts about their knowledge and predictions regarding whether lay-people can avoid or mitigate IoT threats and if so, how this can be achieved. Third, comparing journalists' perceived threat models and individual security behaviours against the expert descriptions of the IoT threat landscape at present and in future, to assess the current and anticipated efficacy of journalists' protective measures, as well as to determine any potential improvements. 

Therefore, given the heightened risk that journalists and sources face from information threats, our surveyed experts responses suggest that most mitigation techniques for current and anticipated IoT threat models will only work in the short-term. The experts recommended that the public pursue large-scale structural responses to the predicted IoT market boom. Again, although the results of our exploratory study cannot be seen as generalisable, they have corroborated existing research (e.g., \cite{mcgregor2016obs}) which also found that journalists' conceptions of their own digital threat models may not be ideal. 

There is therefore a clear need for future academic investigation into socio-technical and policy proposals to minimise IoT-related threats, particularly to the free press. Additionally, a continued push by the Human-Computer Interaction Security academic community for design requirements to be shifted to incorporate better device security, so that the onus is not on users to be experts in managing ambient security is crucial. This is especially true regarding swathes of investigative journalists, whose work can mean that they have to grapple with targeted physical and legal threats, in addition to the complex combination of emergent threat models presented by the now-ubiquitous IoT. 

\section*{Acknowledgment}
We thank all of the participants of this study for their time and contribution. We are additionally grateful for the contribution of the Association for International Broadcasting, which was integral to this study. Anjuli R. K. Shere would also like to recognise the support and guidance of Miranda R. Melcher, which is invaluable.

\bibliographystyle{IEEEtran}
\bibliography{references.bib}

\begin{thebibliography}{10}
\providecommand{\url}[1]{#1}
\csname url@samestyle\endcsname
\providecommand{\newblock}{\relax}
\providecommand{\bibinfo}[2]{#2}
\providecommand{\BIBentrySTDinterwordspacing}{\spaceskip=0pt\relax}
\providecommand{\BIBentryALTinterwordstretchfactor}{4}
\providecommand{\BIBentryALTinterwordspacing}{\spaceskip=\fontdimen2\font plus
\BIBentryALTinterwordstretchfactor\fontdimen3\font minus
  \fontdimen4\font\relax}
\providecommand{\BIBforeignlanguage}[2]{{%
\expandafter\ifx\csname l@#1\endcsname\relax
\typeout{** WARNING: IEEEtran.bst: No hyphenation pattern has been}%
\typeout{** loaded for the language `#1'. Using the pattern for}%
\typeout{** the default language instead.}%
\else
\language=\csname l@#1\endcsname
\fi
#2}}
\providecommand{\BIBdecl}{\relax}
\BIBdecl

\bibitem{forbes16}
{Forbes}, ``{Roundup Of Internet Of Things Forecasts And Market Estimates},''
  2016, {URL:}
  \url{https://www.forbes.com/sites/louiscolumbus/2016/11/27/roundup-of-internet-of-things-forecasts-and-market-estimates-2016/}.

\bibitem{minerva15}
R.~Minerva, A.~Biru, and D.~Rotondi, ``{Towards a definition of the Internet of
  Things (IoT)},'' 2015, {IEEE}.

\bibitem{BBC19}
{BBC News}, ``Security gadgets ``making people more vulnerable'','' 2019,
  {URL:}
  \url{https://www.bbc.com/news/av/uk-england-beds-bucks-herts-49801267/security-gadgets-making-people-more-vulnerable-from-hackers}.

\bibitem{DCMS18}
{UK DCMS}, ``Smart devices: using them safely in your home,'' 2018, {URL:}
  \url{https://www.gov.uk/government/publications/smart-devices-using-them-safely-in-your-home/smart-devices-using-them-safely-in-your-home}.

\bibitem{BBC20}
{BBC News}, ``{Government plans new laws for smart gadgets sold in UK},'' 2020,
  {URL:} \url{https://www.bbc.co.uk/news/technology-51271717}.

\bibitem{alladi2020consumer}
T.~Alladi, V.~Chamola, B.~Sikdar, and K.-K.~R. Choo, ``{Consumer IoT: Security
  vulnerability case studies and solutions},'' \emph{IEEE Consumer Electronics
  Magazine}, vol.~9, no.~2, pp. 17--25, 2020.

\bibitem{hsu2016empirical}
C.-L. Hsu and J.~C.-C. Lin, ``An empirical examination of consumer adoption of
  internet of things services: Network externalities and concern for
  information privacy perspectives,'' \emph{Computers in Human Behavior},
  vol.~62, pp. 516--527, 2016.

\bibitem{hassan2019current}
M.~b. Mohamad~Noor and W.~H. Hassan, ``{Current research on Internet of Things
  (IoT) security: A survey},'' \emph{Computer Networks}, vol. 148, pp.
  283--294, 2019.

\bibitem{nurse2017security}
J.~R.~C. Nurse, S.~Creese, and D.~De~Roure, ``Security risk assessment in
  internet of things systems,'' \emph{IT professional}, vol.~19, no.~5, pp.
  20--26, 2017.

\bibitem{roryp19}
{Rory Peck Trust}, ``{Digital Risk Assessment},'' 2019, {URL:}
  \url{https://rorypecktrust.org/freelance-resources/digital-security/digital-risk-assessment/}.

\bibitem{unrwb17}
{UNESCO and Reporters without borders}, ``{Safety guide for journalists: a
  handbook for reporters in high-risk environments},'' 2017, {URL:}
  \url{https://unesdoc.unesco.org/ark:/48223/pf0000243986}.

\bibitem{ellao19}
J.~A.~J. Ellao, ``{What you need to know about the ongoing cyber-attacks vs.
  alternative news Bulatlat},'' 2019, {URL:}
  https://www.bulatlat.com/2019/02/07/what-you-need-to-know-about-the-ongoing-cyber-attacks-vs
  -alternative-news-bulatlat/.

\bibitem{mindanews19}
{MindaNews}, ``{STATEMENT: DDoS attacks on NUJP, alternative media continue},''
  2019, {URL:}
  \url{https://www.mindanews.com/statements/2019/02/statement-ddos-attacks-on-nujp-alternative-media-continue/}.

\bibitem{mcgregor2015investigating}
S.~E. McGregor, P.~Charters, T.~Holliday, and F.~Roesner, ``Investigating the
  computer security practices and needs of journalists,'' in \emph{Proceedings
  of the 24th USENIX Conference on Security Symposium}, 2015, pp. 399--414.

\bibitem{mcgregor2016individual}
S.~E. McGregor, F.~Roesner, and K.~Caine, ``Individual versus organizational
  computer security and privacy concerns in journalism,'' \emph{Proceedings on
  Privacy Enhancing Technologies}, vol.~4, pp. 418--435, 2016.

\bibitem{mcgregor2017weakest}
S.~E. McGregor, E.~A. Watkins, M.~N. Al-Ameen, K.~Caine, and F.~Roesner, ``When
  the weakest link is strong: secure collaboration in the case of the panama
  papers,'' in \emph{Proceedings of the 26th USENIX Conference on Security
  Symposium}, 2017, pp. 505--522.

\bibitem{henrichsen2019breaking}
J.~R. Henrichsen, ``Breaking through the ambivalence: Journalistic responses to
  information security technologies,'' \emph{Digital Journalism}, pp. 1--19,
  2019.

\bibitem{erlingsson2017hands}
C.~Erlingsson and P.~Brysiewicz, ``A hands-on guide to doing content
  analysis,'' \emph{African Journal of Emergency Medicine}, vol.~7, no.~3, pp.
  93--99, 2017.

\bibitem{williams2017privacy}
M.~Williams, J.~R.~C. Nurse, and S.~Creese, ``Privacy is the boring bit: user
  perceptions and behaviour in the internet-of-things,'' in \emph{15th Annual
  Conference on Privacy, Security and Trust (PST)}.\hskip 1em plus 0.5em minus
  0.4em\relax IEEE, 2017, pp. 181--18\,109.

\bibitem{theguard19}
{The Guardian}, ``{Australia's anti-encryption laws being used to bypass
  journalist protections, expert says},'' 2019, {URL:}
  https://www.theguardian.com/australia-news/2019/jul/08/australias-anti-encryption-laws-being-used-to-bypass-journalist-
  protections-expert-says.

\bibitem{privint19}
{Privacy International}, ``{Case Study: Connected Cars and the Future of Car
  Travel in the Digital Age},'' 2019, {URL:}
  \url{http://privacyinternational.org/case-studies/769/case-study-connected-cars-and-future-car-travel-digital-age}.

\bibitem{conti2018internet}
M.~Conti, A.~Dehghantanha, K.~Franke, and S.~Watson, ``{Internet of Things
  security and forensics: Challenges and opportunities},'' \emph{Future
  Generation Computer Systems}, vol.~78, no.~2, pp. 544--546, 2018.

\bibitem{vicari19jot}
J.~Vicari, ``{Journalism-of-Things Conference 2019. We’ll just say it:
  Journalism and IoT are a perfect match},'' 2019, {URL:}
  https://medium.com/journalism-of-things/well-just-say-it-journalism-and-iot-is-a-perfect-
  match-ca3d7bbdbb04.

\bibitem{mills2016reluctant}
A.~Mills and K.~Sarikakis, ``{Reluctant activists? The impact of legislative
  and structural attempts of surveillance on investigative journalism},''
  \emph{Big Data \& Society}, vol.~3, no.~2, pp. 1--11, 2016.

\bibitem{theguar19teen}
{The Guardian}, ``{Teen claims to tweet from her smart fridge – but did she
  really?}'' 2019, {URL:}
  \url{https://www.theguardian.com/technology/2019/aug/13/teen-smart-fridge-twitter-grounded}.

\bibitem{nurse2015smart}
J.~R.~C. Nurse, A.~Erola, I.~Agrafiotis, M.~Goldsmith, and S.~Creese, ``Smart
  insiders: exploring the threat from insiders using the internet-of-things,''
  in \emph{2015 International Workshop on Secure Internet of Things
  (SIoT)}.\hskip 1em plus 0.5em minus 0.4em\relax IEEE, 2015, pp. 5--14.

\bibitem{maiti2019light}
A.~Maiti and M.~Jadliwala, ``Light ears: Information leakage via smart
  lights,'' \emph{Proceedings of the ACM on Interactive, Mobile, Wearable and
  Ubiquitous Technologies}, vol.~3, no.~3, pp. 1--27, 2019.

\bibitem{boingb16}
{Boing Boing}, ``{A lightbulb worm could take over every smart light in a city
  in minutes},'' 2016, {URL:}
  \url{https://boingboing.net/2016/11/09/a-lightbulb-worm-could-take-ov.html}.

\bibitem{dcms19sbd}
{UK DCMS}, ``{Secure by Design},'' 2019, {URL:}
  \url{https://www.gov.uk/government/collections/secure-by-design}.

\bibitem{etsi19iot}
{ETSI}, ``{Cyber Security for Consumer Internet of Things},'' 2019, {URL:}
  \url{https://www.etsi.org/deliver/etsi_ts/103600_103699/103645/01.01.01_60/ts_103645v010101p.pdf}.

\bibitem{theecon19}
{The Economist}, ``{Britain is rejecting an ever greater share of
  freedom-of-information requests},'' 2019, {URL:}
  https://www.economist.com/graphic-detail/2019/05/01/britain-is-rejecting-an-ever-greater-share-of-freedom-of-
  information-requests.

\bibitem{mcgregor2016obs}
S.~E. McGregor, P.~Charters, T.~Holliday, and F.~Roesner, ```security by
  obscurity': Journalists’ mental models of information security.'' in
  \emph{Proceedings of the International Symposium on Online Journalism}, 2016.

\end{thebibliography}

\newpage
\appendices

\section{Interview Questions}
\noindent \textit{The aim of this interview is to gain insight into your cyber security perceptions and practices as a member of the media industry. }
\begin{enumerate}
\item In order to help you understand the equipment and systems we are researching which of these devices do you use?
\begin{itemize}
  \item Laptop/desktop computer
  \item Smart phone/tablet
  \item Smart TV
  \item Voice operated assistant, i.e. Alexa, Google Home etc.
  \item Wearable smart technology, e.g. fitness trackers, smart watches
  \item Cloud Storage
  \item Location tracking devices, i.e. sat nav, ``find my phone'', ``find my keys''
  \item Smart phone-enabled locks, i.e. remote-access door locks
  \item Smart toys, e.g. gaming systems, children's toys, recreational use drones
  \item Internet-connected cars
  \item Internet-connected security systems
  \item Internet-connected HVAC and lighting systems.
\end{itemize}

\textit{Let us know look at your general journalism experience. }
\item How long have you been a journalist?
\item What motivated you to become a journalist?
\item Do you specialise in any particular journalism area and within your career what general areas of journalism have you covered? 
\item What motivated you to specialise in this/these areas?
\item Have you worked in any other jobs other than journalism within your career?\\

\textit{Let us now turn to cyber security and the risks to journalists from the IoT and more generally the ``online'' world we all now inhabit.}
\item Is cyber security something that you consider while carrying out your day-to-day work?
\item Do you think that cyber security is relevant to your work as a journalist? If so, how?
\item Is cyber security a subject element of your journalism?
\item Have you been the target of any digital attacks? If yes, please give details
\item What distinctions, if any, do you perceive between the following four categories:
\begin{enumerate}
 \item ``internet-enabled devices'',
 \item ``internet-connected devices'',
 \item ``Internet of Things devices'' or 
 \item ``smart devices''. 
\end{enumerate}

\textit{In this interview, the phrase ``Internet of Things'' will be used to refer to ``smart (i.e. analytically capable), internet-connected devices that can share data with each other, creating a `network' of devices''. }
\item Do you have any concerns about these sorts of devices?
\item Are there any of these devices you would not use and for what reasons?
\item Do you have any privacy concerns about any of these devices?
\item Would you say your use/views of these devices are typical of journalists you know?
\item What knowledge do you have of cyber-attacks on journalists' devices? Or alternately, what knowledge do you have of cyber-attacks on journalists?

\textit{Let us look at what you think about the people who are attacking journalists. }
\item Who would you consider being the perpetrators of these threats?
\item What do you think are the objectives of cyber-threat actors who are targeting journalists?
\item How at risk do you feel, as a journalist, with respect to exploitation of new technologies like Internet of Things devices?
\item How do you think that threats against journalists and their sources have evolved with the introduction and expansion of Internet of Things devices?
\item To what extent do you perceive threats to you from or via Internet of Things devices?\\

\textit{Let us now look at threats that are not cyber based.}
\item As a journalist, did you experience threats before Internet of Things devices became very popular? What sort of other threats did you experience?
\item What other threats were you aware of?
\item Are you concerned with the IoT's increased potential for attacks to jump from the cyber realm into the physical world?
\item Are you aware of any combined threats that involve cyber and non-cyber-attacks?\\

\textit{Let us now move on to your personal cyber security strategies.}
\item Would you say you have personal cyber security strategies?
\item What are your strategies or tactics for protection of your Internet of Things devices against cyber-attacks?/ What are your strategies for defence against IoT-related threats?
\item Are these specific to your profession or generic, and in what way are they specific if they are?
\item Has your approach to source protection changed in light of your understanding of new and evolving cyber-threats, specifically related to Internet of Things devices? 
\item Do you warn your sources about Internet of Things device-related risk to their anonymity and security?\\

\textit{There have been many examples of journalists being subject to surveillance, censorship and intimidation specifically by governments and law enforcement agencies. Sometimes what would be discerned as friendly states. These state players are making greater use of internet-enabled device exploitation. So, let's now look at governments and law enforcement agencies' Internet of Things exploitative capabilities as they relate to you and your work. We are not looking for general or rumoured issues but ones you have personally encountered, and we understand the potential sensitivity of this so it will be kept highly confidential.  }
\item Given any sensitive topics that you may have covered, do you feel there is a threat to you from any country's governments or law enforcement agencies?
\item What do you believe the threat is, can you describe it?
\item If you are willing to say, which countries and which agencies?
\item If you are not willing, can you say why you are not willing to identify them?\\

\textit{Looking to the future for journalists.}
\item From your knowledge of the industry, do you think that journalists are becoming more aware of cyber-threats and effectively protecting themselves?
\item Whose responsibility is it to improve journalists' understanding of threats through IoT devices? (i.e. the journalists themselves, device suppliers, media organisations, etc.) 
\item Do you need better technical systems as well as knowledge? 
\item Is there an aspect of journalists' cyber security that you feel is disproportionately overlooked?
\item Do you have anything you'd like to add or comment on?
\end{enumerate}

\section{Cyber Security Expert Survey}
Please rate the level of your knowledge about cyber security: \textit{[Likert Scale with 5 levels based on knowledge]}
\begin{enumerate}
\item What do you think are the digital threats associated with the use of smart devices and the Internet of Things? \textit{[Free text box]}
\item What do you think are the physical threats associated with the use of smart devices and the Internet of Things? \textit{[Free text box]}
\item To what extent do you think members of the general public are able to opt-out of their use of smart devices and the Internet of Things now? \textit{[Can opt-out/Can't opt-out]}
\item To what extent do you think members of the general public will be able to opt-out of their use of smart devices and the Internet of Things in the future (e.g. over the next 5 years)? \textit{[Will be able to opt-out/Won't be able to opt-out]}
\item What do you think are some of the ways in which members of the public are able to mitigate digital and physical threats associated with the use of smart devices and the Internet of Things? \textit{[Free text box]}
\end{enumerate}

\end{document}